\documentclass[%
 reprint,
 superscriptaddress,
 amsmath,amssymb,
 aps,
 pre,
 floatfix,
]{revtex4-1}

\usepackage{graphicx}% Include figure files
\usepackage{dcolumn}% Align table columns on decimal point
\usepackage{bm}% bold math
\usepackage{natbib}
\usepackage{physics} % bra & ket notations as well as Trace notations 
\usepackage{ bbold }
\usepackage[utf8x]{inputenc}
\usepackage{hyperref}% add hypertext capabilities
\usepackage[mathlines]{lineno}% Enable numbering of text and display math
\usepackage{color}
\usepackage{amsthm}
\usepackage{amsmath}
\usepackage{amssymb}
\usepackage{mathrsfs}
\usepackage{array}
\usepackage{graphicx}
\usepackage[normalem]{ulem}
\usepackage{verbatim,bm}
\usepackage[normalem]{ulem}
\usepackage{enumerate}% roman numerals for itemize

\usepackage{multirow}

\newcommand{\red}[1]{\textcolor{black}{#1}} % NEW

\usepackage{hyperref}
\begin{document}

% \preprint{APS/123-QED}

\title{ Power law decay of entanglement quantifiers in a single agent to a many body system coupling    }% Force line breaks with \\
%\thanks{}%

\author{Ohad Shpielberg}
\email{ohads@sci.haifa.ac.il}
% \email{ohad.shpilberg@college-de-france.fr}
\affiliation{ Haifa Research Center for Theoretical Physics and Astrophysics, University of Haifa, Mt. Carmel, Haifa 31905, Israel}
%\affiliation{ Collège de France, 11 place Marcelin Berthelot, 75231 Paris Cedex 05 - France}

%\collaboration{}%\noaffiliation

\date{\today}

\begin{abstract}

%Control over a many body quantum system can be achieved by entangling the system to a single agent.  Here, we study the ground state entanglement properties of two site lattice models, between a highly occupied subsystem (site $A$) to a few particles (site $B$). Both the Von Neumann entanglement entropy and the Logarithmic negativity show a power law decay in $R$ --  the occupancy ratio between the systems.
%This implies that it is feasible to entangle large many body systems to a single atom, as recently reported experimentally.         

\red{
Manipulating many body quantum systems is a challenge. A useful way to achieve it would be to entangle the system to a diluted system, with a small particle number.  Preparation of such entangled states can be facilitated as ground state of a many body Hamiltonian or the steady state of a many body open quantum system.
Here we study two-site lattice models with a conserved boson number,  biased to display a large occupancy in one of the sites. The Von Neumann entanglement entropy as well as the Logarithmic negativity show a typical power law decay in  $R$, the occupancy ratio between the two sites. These results imply that it is feasible to entangle a large many body system to a single atom, as recently reported experimentally. 
}

\end{abstract} 

\maketitle

\section{Introduction}

% \begin{itemize}
%     \item Step 1: 
%     \begin{itemize}
%     \item Move 1: Entanglement is a key concept in quantum information, quantum metrology etc.
%     \item Move 2 and 3: How to entangle systems (steady state Hamilonian / Lindblad), how to measure entanglement experimentally (entanglement witnesses \& entanglement measures). 
% \end{itemize}
% \item Step 2 (Move 2): In 1-3000, it was experimentally demonstrated that a single photon can be entangled with a MB system of 3000 atoms. Measurement of the photon allows to infer \& control the MB state of the system.  Unlike the case of extended system where the area law generally rules, a general rule for the bipartite entanglement between a single agent to a MB system is unknown.  
% \item Step 3: showing that the niche can be occupied. 
%     \begin{itemize}
%         \item Step1: the purpose of this research is to study ... 
%         \item We find that  characteristically $E(R) ~ 1/ R^\alpha $ for $\alpha \geq 0$. There can be logarithmic correction to this power law. In the large $R$ limit, this is of little consequence. 
%         \item The structure of this work is as follows 
%     \end{itemize}
% \end{itemize}

Entanglement is a key resource in quantum information \cite{Entanglement_Review_Horodecki,Entanglement_MB_Vedral,Vedral_Book}, quantum computing \cite{NielsenChuang,preskill2018quantum} and quantum metrology \cite{MetrologyTreutlein_Review}. Recently, there has been significant advancement in generating, manipulating and measuring entangled many body states; both experimentally and theoretically \cite{omran2019generation,islam2015measuring,kaufman2016quantum,kallush2021controlling}.  
%A common starting point is to study bipartite entanglement, where the system $AB$  is partitioned resulting in the subsystems $A$ and $B$. 
Both preparation of the entangled state and its validation using, e.g. entanglement witnesses \cite{Witnesses_Gniewomir}, are challenging aspects in many body systems and are the focus of ongoing research \cite{lacroix2020symmetry,weimer2010rydberg,carr2013preparation,VerstraeteWolfCirac}.

\begin{figure}
    \centering
    \includegraphics[scale=0.45]{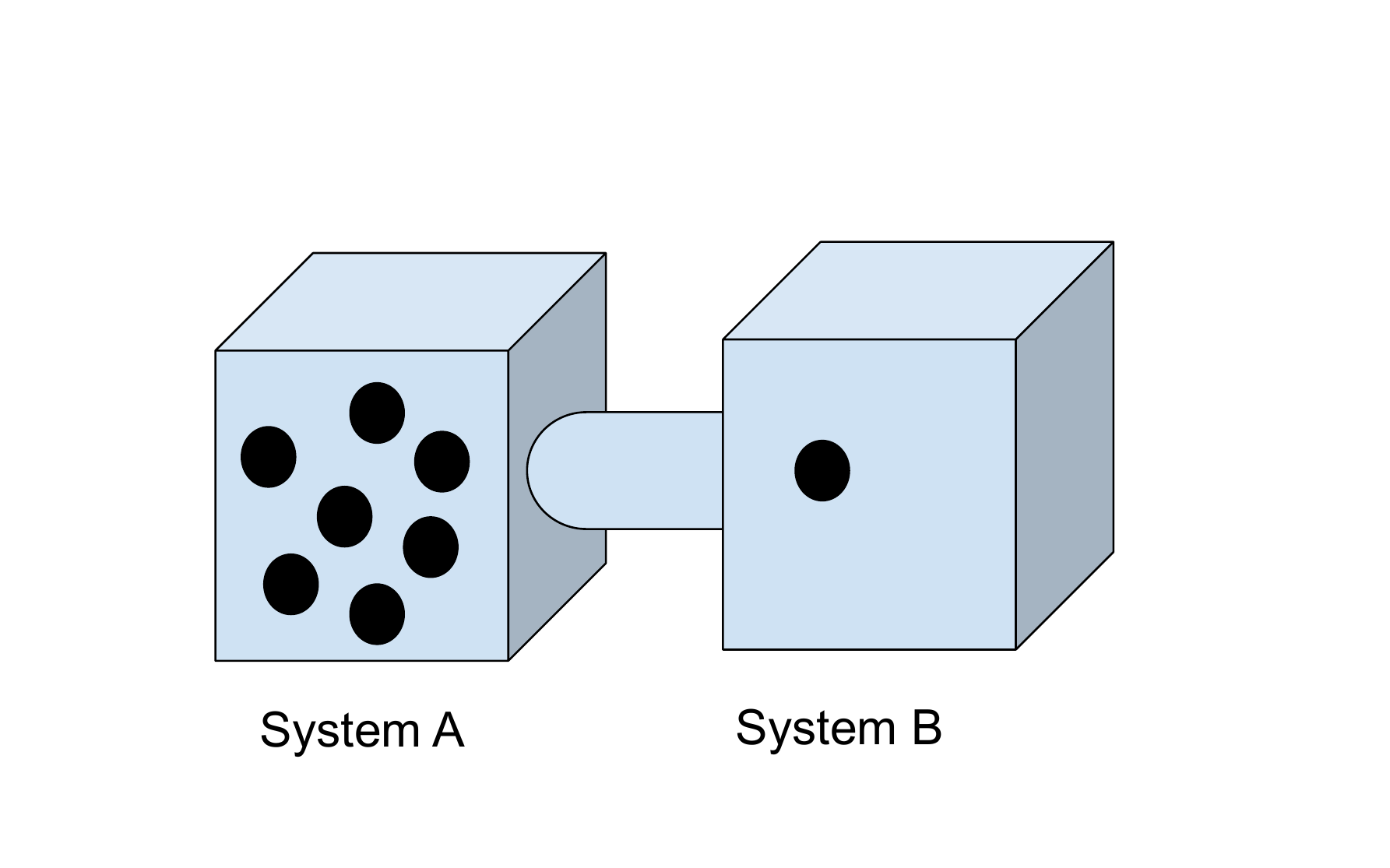}
    \caption{The systems $A$ and $B$ are made to interact such that we find a large occupancy of particles at system $A$ and a low occupancy at system $B$.  }
    \label{fig:ABsystem}
\end{figure}

Preparation of a desired entangled state can be realized as the ground state of a carefully designed Hamiltonian. Therefore, understanding the entanglement properties of ground states is of practical importance. Significant effort has been directed for extended systems, especially in $1D$. The ground state of a Hamiltonian with local interactions typically exhibits an area law in the bipartite Von Neumann entanglement entropy \cite{Eisert_AreaLaw}, contrary to the generic volume law of typical quantum states.  However, the area law does not characterise a system composed of a few sites, where each site can occupy a large number of particles (see Fig.~\ref{fig:ABsystem}). For two sites, the average Von Neumann entanglement entropy is known \cite{Page_averageEnropy,sen1996average}, \red{but the characteristic properties of the ground state entanglement remain largely unexplored.} %but the ground state entanglement properties are not. 

A particularly appealing case is for a single agent in system $B$ to be entangled to a large number of particles in system $A$. Such entanglement allows to manipulate the many body system via the single agent. This setup was experimentally demonstrated in \cite{McConnell2015}, where a single photon was entangled with roughly $3,000$ atoms. At this point, it is unknown whether there is a limit to the number of particles that could realistically be entangled with a single agent. It is further unknown whether our setup leads to typical ground state entanglement properties.

To answer the above questions, the entanglement needs to be quantified \cite{Huber_EntanglementCost_Review,Plenio_Entanglement_Review}. Choosing an appropriate entanglement quantifier depends on the intended application, e.g. entanglement distillation to produce  Bell states. However, the entanglement quantifier can alternatively be chosen to accommodate fast calculations of known density matrices, e.g. the Logarithmic negativity $E_{ln}$. For pure states, the Von Neumann entanglement entropy $E_{vn}$ serves both purposes.

\red{Keeping in mind the motivation of entangling a single agent to a many body system, we turn to a simpler theoretical setup. } In this work we consider $N $  bosons, occupying a two-site system. We study the ground state entanglement properties when the \red{system is tuned to display an} overwhelming majority of bosons \red{occupying} site $A$ (see Fig.~\ref{fig:ABsystem}). To make this statement precise, let $\hat{n}_{A,B}$ be the corresponding number operators. Then, $R = \langle \hat{n}_A \rangle / \langle \hat{n}_B \rangle $ is the ratio of the particle occupancies.  We study the large $R$ behaviour of both the ground state Von Neumann entanglement entropy  and Logarithmic negativity of the Bose-Hubbard Hamiltonian. \red{For open quantum systems, the same setup can be considered, where the steady state takes the role of the ground state.} %Then, w
We study the steady state Logarithmic negativity of a Lindblad super-operator model -- the quantum asymmetric inclusion process at large $R$ values \cite{eisler2011crossover,Bernard18}. 

Both models are studied at different scaling regimes. Nevertheless, they all  consistently lead to a power law decay in the entanglement quantifiers. Quantitatively, the Von Neumann entanglement entropy $E_{vn} \sim \frac{\log R}{R^\alpha} $ and the Logarithmic negativity  $E_{ln} \sim \frac{1}{R^\alpha}$ for $R\gg 1$. See Tables \ref{table:BH model table} and \ref{table:QASIP model table} for a summary of the results. 

\red{We argue that the power law decay is typical,}
%The power law decay seem to be typical 
as we have considered  two disparate models and different scaling regimes for each model. The slow power law decay, contrasting with an exponential decay,  \red{answers in a quantifiable way how realistic it is} to entangle a single\red{, or a few} atoms to a highly occupied many body state. The exponent $\alpha$ is non-universal. Therefore, interacting systems that result in small $\alpha $ values are favorable to facilitate entanglement between the \red{diluted system} to the large occupancy system.

The structure of this paper is as follows. In Sec.~\ref{sec:Results} we present the Hamiltonian and Lindblad models and summarise the main results. % In Sec.~\ref{sec:Entanglement}  we recall the definition and usefulness of the entanglement quantifiers. 
Sec.~\ref{sec:Analytical treatment} presents in full the analytical and numerical treatment of the systems under study.  Finally,  Sec.~\ref{sec:Discussion}  recaps the main findings and their physical relevance and suggests future directions. 

%%% ==================================== 
\section{Models and results \label{sec:Results}}

% \begin{itemize}
%     \item Lattice models are natural to consider as manifestations of the composite system setup. In particular, we study lattice models with two sites, $A$ and $B$.
%     \item First, we study the entanglement properties at large $R$ of the ground state of the BH Hamiltonian with $N$ conserved bosons.  
%     \item Explain BH physics 
%     \item state the results and summarise in table. \item Then, we consider the Lindblad dynamics corresponding to a quantum analog of the asymmetric exclusion process with . 
%     \item Explain QASIP physics 
%     \item state the results and summarise in table.
%     \item Thirdly, we consider continuum systems, where there is no clear distinction between the subsystems. As a simple example, we take the ground state of a BEC with $N$ particles. One can fictitiously partition the system and show that the ground state is entangled. However, for large $R$ values corresponding to the box $B$ to be of a finite fraction of the total volume of the BEC, the entanglement saturates to a constant (with 1/R corrections). \blue{ explanation in discussion: The continuum limit leads to loss of information and thus a constant term arises in the entanglement entropy. The power law subleading corrections are probably the echo, remaining from the discretized case.  }
% \end{itemize}

The aim of this work is to quantify the bipartite entanglement of a composite $AB$ system at large $R$ values. Therefore, it is natural to study lattice models, where the distinction between the two subsystems is clear cut. In particular, we study lattice models with two sites, $A$ and $B$.  

\red{To demonstrate the power law behaviour is typical,} two disparate lattice models are considered. First, the ground state entanglement of the \red{two-site} Bose-Hubbard model is extensively studied. Second, we consider a generalization of the asymmetric inclusion process \cite{Grosskinsky_ASIP} to the quantum realm via a Lindblad equation, dubbed here the quantum asymmetric inclusion process (QASIP). We then study the entanglement properties of the steady state \red{at large $R$}. 

%A continuum model of a BEC ground state is also considered. There, the typical behaviour is shown to saturate for large $R$ values. 

% Here we study two lattice models of two sites; The Bose-Hubbard model and a Lindblad type open quantum system which can be interpreted as a quantum version of the asymmetric inclusion process. \blue{ref to the relevant literature }. We show that in these two setups, the entanglement quantifiers in question, i.e. the logarithmic negativity and when applicable the Von Neumann entanglement exhibit a power law behaviour at large $R$ values. See tables \ref{table:BH model table}  and \ref{table:QASIP model table} for the leading order behaviour.  We stress that the power law behaviour is attained for different scaling limits and for two independent physical models. Therefore, the power law behaviour is conjectured to be general for lattice systems.    

% We also consider a continuum model associated with a Bose-Einstein condensation. The subsystems $A,B$ are arbitrarily determined and result in the entanglement quantifiers saturating at a constant value for large $R$. The violation of the power law behaviour in continuum systems may suggest that it is coarse graining of lattice models that leads to the entanglement power law violation. 

% In the rest of this section, we present the three models. Furthermore, we motivate the scaling limits leading to large $R$ behaviour and state the results.

%============================================================================================== %

\subsection{The \red{two-site} Bose-Hubbard model}

The Bose-Hubbard model is a simple yet rich many body lattice model of spin-less bosons. It allows studying the superfluid-insulator transition \cite{Fisher_BoseHubbard_SFinsulator} and can be experimentally implemented using optical lattices \cite{jaksch1998cold,greiner2002quantum}.  
\red{The particular case of the two-site Bose-Hubbard model was extensively used in the literature to study tunneling effects between potential wells \cite{LinksTwositeBH} as well as fragmentation \cite{Sippe_fragmentation}.   Importantly for our purposes,} the two site Bose-Hubbard model is expected to be both analytically tractable and to present typical physical behaviour in terms of the ground state entanglement. Hence, it serves as the starting point of our analysis.

The \red{two-site} Bose-Hubbard Hamiltonian is given by 
\begin{equation}
    \label{eq:BH Hamiltonian}
    H_{BH} = -J  (\hat{b}^\dagger _A \hat{b}_B  + \hat{b}^\dagger _B \hat{b}_A  ) - \mu  \hat{n}_A + \frac{U}{2} \sum_{i=A,B} \hat{n}_i- \hat{n}^2 _i,
\end{equation}
where  
$J$ is the hopping matrix element between neighboring sites and $U$ determines the strength of the on-site interaction. The operators $\hat{b}_i,\hat{b}^\dagger _i $ are the site-dependent bosonic creation and  annihilation operators and $\hat{n}_i = \hat{b}^\dagger _i \hat{b} _i  $ is the number operator for $i=A,B$. In \red{an} optical lattice, the potential wells are represented by the two sites \cite{jaksch1998cold}. The potential offset between the two \red{asymmetric} potential wells is given by $\mu$. It furthermore allows to imbalance the system towards large $R$ values.

It is useful to note that the total particle number, $ \hat{N} =   \hat{n}_A + \hat{n}_B   $ is conserved. Therefore, we analyze the ground state with $N$ bosons. In what follows, we consider two scaling schemes leading to large $R$ values. 

\begin{figure}
    \centering
    \includegraphics[scale=0.37]{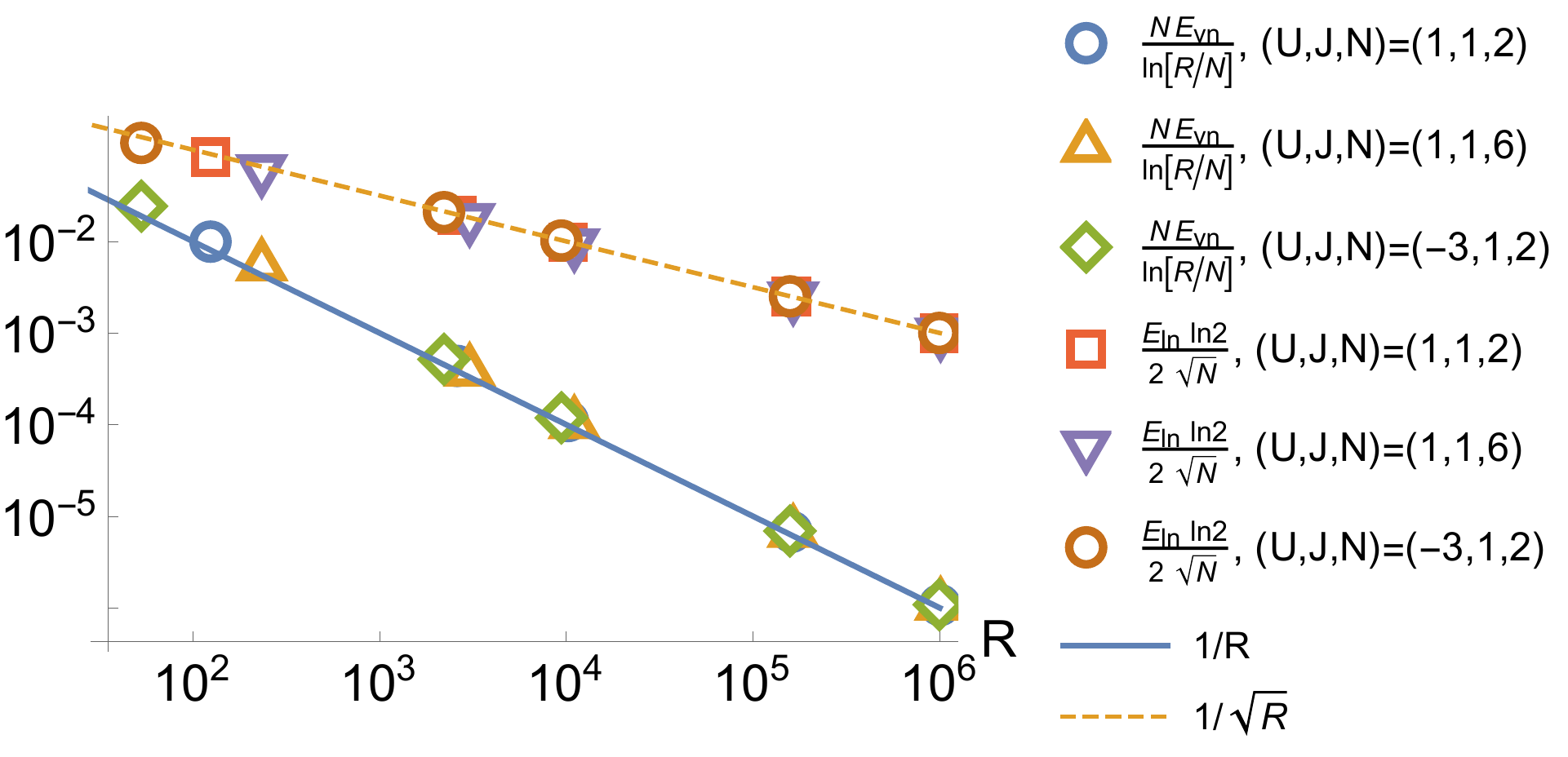}
    \caption{
    \red{Corroborating the perturbation theory analytical predictions presented in Table~\ref{table:BH model table}. The entanglement quantifiers are evaluated numerically for the two-site Bose-Hubbard model and compared  with perturbation theory. Different  $U,J,N$ values are considered (see legend), in the range $\mu \in  \left[10,10^3\right]$ to facilitate large $R$ values. No fitting parameters are required to observe the collapse onto the expected power law behaviour.    }
    } 
    \label{fig:BH large mu}
\end{figure}

First, taking large $\mu$ values and keeping $N,J$ and $U$ fixed,
\red{a perturbative treatment leads to} 
$R = \mu^2 /J^2 + O(\mu) $. \red{See Sec.~\ref{sec:Analytical treatment}}. In this limit, and as long as $\frac{J \sqrt{N}}{  \mu } , \frac{U N^2}{ \mu}   \ll 1 $, we find analytically the power law behaviour described in Table~\ref{table:BH model table}. These results are also numerically corroborated in Fig.~\ref{fig:BH large mu}. % \ref{fig:Evn Eln comparison}.
Note that the logarithmic correction in the Von Neumann entanglement  hardly changes the behaviour from a clean power law.

\begin{figure}
    \centering
    \includegraphics[scale=0.38]{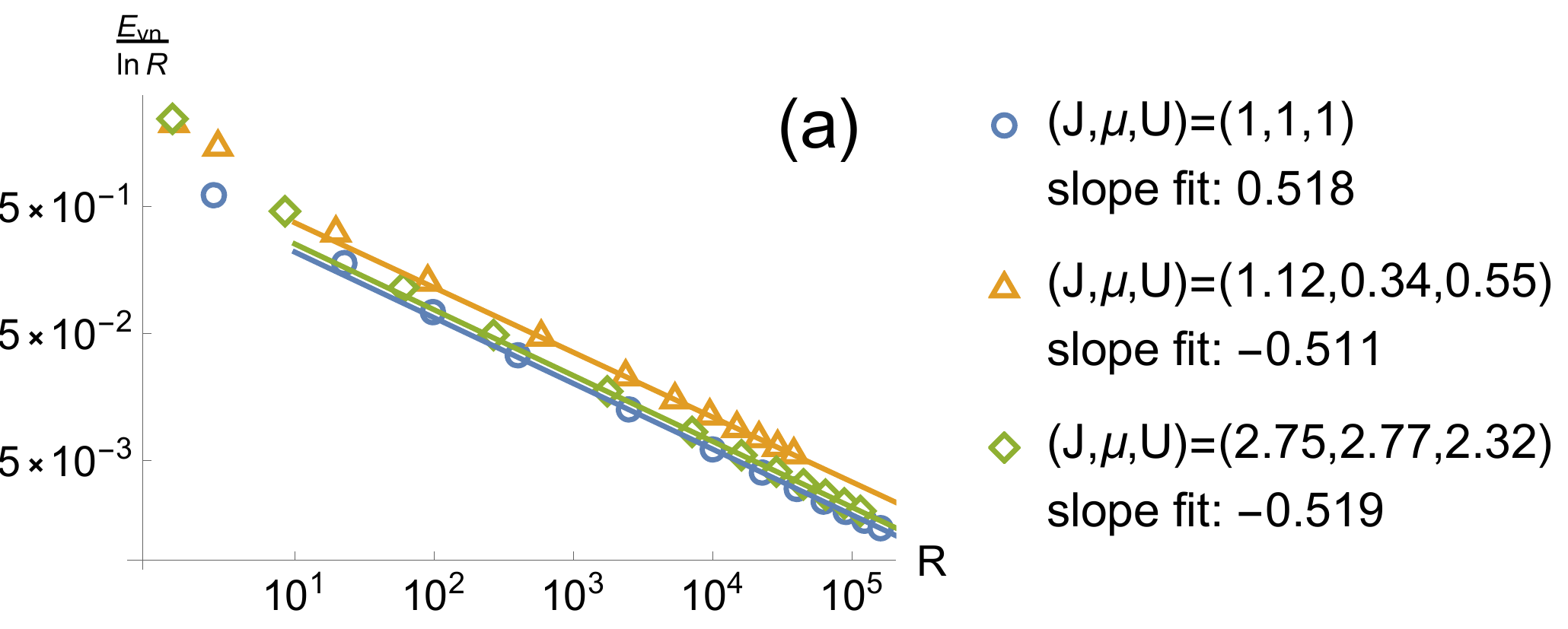}
    \includegraphics[scale=0.38]{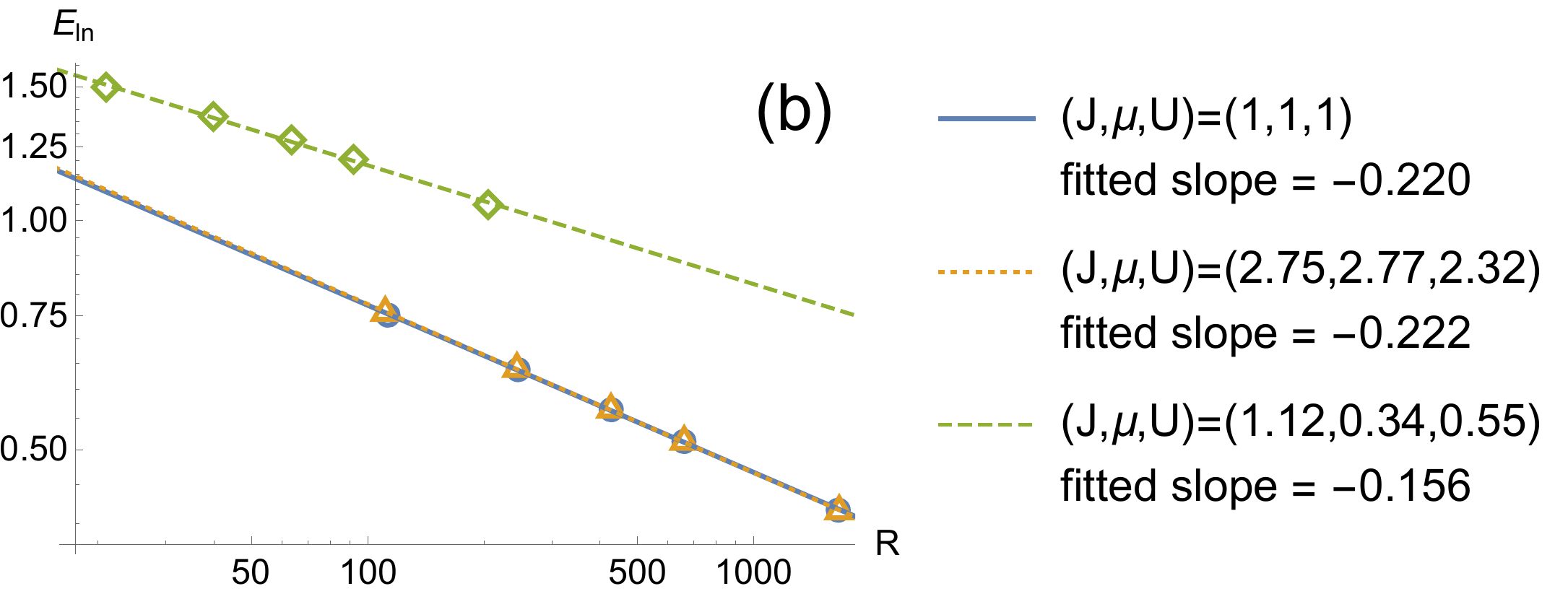}
    \caption{
    \red{Numerical fitting of the entanglement quantifiers at the large $N$ limit of the two-site Bose-Hubbard model. Both $\frac{E_{vn}}{\log R}$ and $E_{ln}$ exhibit power law behaviour at large $R$ values, albeit with different exponents. Numerical convergence of  arbitrary parameters to the same exponent is obtained when sufficiently large $R$ values are reached. 
    (a) The Von Neumann entanglement entropy is shown to scale like $E_{vn} \propto \frac{\ln R}{R^\alpha}$, where $\alpha \approx 0.518$. The range $N \in  \left[10,400 \right]$ was tested.
    (b) The logarithmic negativity is shown to scale like $E_{ln} \propto R^{-\alpha}$, where $\alpha \approx 0.22$ is numerically fitted in the range $N \in  \left[10,40 \right]$. See the Appendix \ref{app:BH model} for a detailed discussion on the reduced range of $N$ for the logarithmic negativity.  }  
    }
    \label{fig:BH large N fitting}
\end{figure}

Second, we consider the large $N$ limit, with fixed $\mu,J$ and $U$. 
A perturbative approach is harder in this case \red{as the Hilbert space of the effective Hamiltonian depends on the particle number $N$ (see Sec.~\ref{sec:Analytical treatment}).}  See Fig.~\ref{fig:BH large N fitting},  Table~\ref{table:BH model table} \red{and the appendix \ref{app:BH model}} for the numerical analysis of the large $N$ limit.

%\begin{tabular}{ |p{1.2cm}||p{2.2cm}|p{2.2cm}|  }
\begin{table}[ht]
\centering
\begin{tabular}{ |c||c|c|c| }
 \hline
  &&&\\ 
  & $R$ & $E_{vn} = \Gamma \frac{\ln R}{   R^{\alpha }}$ & $E_{ln} = \frac{\Gamma}{   R^{\alpha }}$ \\
  &&&\\ 
 \hline
 &&&\\ 
 $\mu \gg 1$   & $\mu^2/J^2 $    & 
 $\Gamma = N , \alpha =1 $ 
 %$\frac{\log R/N}{R/N} $ 
 & $\Gamma = \frac{2 \sqrt{N}}{\ln 2} , \alpha =\frac{1}{2} $
 %$ \frac{2}{\log 2} \sqrt{\frac{N}{R}}$ 
 \\
 &&& \\ 
 $N \gg 1$ &$ \propto  N^{\beta}, \,  \beta \approx 2.15$  & 
 $\alpha \approx 0.518 $
 %$\propto \frac{\log R}{R^{0.55}}  $ 
 & 
 $\alpha \approx 0.220 $ \\
 \hline
\end{tabular}
\caption{The power law behaviour of the ground state entanglement of the Bose-Hubbard model at large $R$ values. \red{For the large $N$ limit, the exponents are evaluated numerically.}}
\label{table:BH model table}
\end{table}

In conclusion, the ground state of the two-site Bose-Hubbard model leads to a power law behaviour of both the Von Neumann entanglement entropy and the Logarithmic negativity at large $R$ values in two different scaling schemes.

Next, we explore the large $R$ \red{steady state} entanglement in an open quantum system setup: the quantum asymmetric inclusion process. 

\subsection{Open quantum system }
% A system of two sites is occupied with $N$ bosons that interact with a set of environments. The dynamics of this open quantum system is given by the GSKL equation ....
% To find the steady state, we need to solve a linear set of $N^2$ equations (notice the decrease from $N^4$ due to conserved quantities of the GKSL equation ). For small values of $N$, one can find a the steady state analytically and for intermediate values it is possible to solve the equations numerically.   
% Since the system yields a mixed density matrix at the steady state, we quantify the entanglement via the Logarithmic negativity. It is plotted as a function of $R$ and can be shown to decay like a power law with the exponent depending on the value of $\gamma,\varepsilon,\eta$. In particular for $\gamma=4,\eta=1,\varepsilon=1$, we find $\alpha \approx 0.23$ (\red{reference to plot}).    

Realistically, physical systems are never truly isolated. Interestingly, the coupling of a quantum system to an environment does not always lead to complete loss of entanglement in the system. Indeed, there are examples where the environment can be engineered to produce a desirable entangled state \cite{VerstraeteWolfCirac,shpielberg2020diffusion}.

Here, we aim to study whether the large $R$ power law behaviour persists for \red{steady states of} open quantum systems as well. To that end, we focus on a quantum analog of the asymmetric inclusion process given in terms of the 
%Gorini–Kossakowski–Sudarshan–Lindblad equation, here the 
Lindblad equation.

 In this setup, one usually assumes that a  quantum system is coupled to an environment with fast relaxation times. This in turn, allows to discard non-Markovian contributions to the evolution of the density matrix and results in the Lindblad equation \cite{BreuerBook,GKS_equation,lindblad1976generators}
\begin{eqnarray}
\label{eq:GKSL equation general}
    \partial_t \rho &=& \mathcal{H}(\rho)  + \sum_k \mathcal{D}_{\hat{L}_k} (\rho)
    \\ \nonumber 
    \mathcal{H}(\rho) &=& -i \left[H,\rho \right]
    \\ \nonumber 
    \mathcal{D}_{\hat{L}} (\rho) &=& \hat{L}\rho  \hat{L}^\dagger - \frac{1}{2} \lbrace \hat{L}^\dagger \hat{L} , \rho \rbrace. 
\end{eqnarray}
In eq.\eqref{eq:GKSL equation general},  $H$ is a Hermitian operator and $\left[ \bullet, \bullet  \right],\lbrace \bullet , \bullet \rbrace$ are the commutation and anti-commutation relations correspondingly. Despite the restriction to Markovian dynamics, enough quantumness remains in the Lindblad equation \cite{Spohn_Lebowitz78,QuantumThermodynamics_Kosloff,Quantum_VDP}. \red{At this point, we turn to study a particular Lindblad model, allowing to facilitate the large $R$ limit analytically.}

The QASIP describes the dynamics of bosons on a two-site lattice, where the boson interactions are environment assisted \cite{eisler2011crossover,Bernard18}.  The evolution of the density matrix is given by 
\begin{align}
\label{eq:Lindblad dynamics}
    \partial_t \hat{\rho} &= 
    \mathcal{L}_{QASIP} (\hat{\rho})
    \\ \nonumber 
    \mathcal{L}_{QASIP} (\hat{\rho})
     & =  
    \mathcal{H}_{\mathrm{tb}}(\hat{\rho}) +  \mathcal{L}_{\mathrm{D}}(\hat{\rho}) + \mathcal{L}_{E}(\hat{\rho})  
    \\ \nonumber 
    H_{\mathrm{tb}}(\hat{\rho}) &= \varepsilon   \left( \hat{b}^\dagger _A \hat{b}_B + \hat{b}^\dagger _B \hat{b}_A \right) 
    \\ \nonumber 
    \mathcal{L}_{\mathrm{D}}(\hat{\rho}) &= \eta   \sum _{k=A,B} \mathcal{D}_{\hat{n}_k} (\hat{\rho})
    \\ \nonumber 
    \mathcal{L}_{\mathrm{E}}(\hat{\rho}) &=  \gamma   \mathcal{D}_{\hat{b}^\dagger _A \hat{b}_B} (\hat{\rho}).
\end{align}
\red{See eq.\eqref{eq:GKSL equation general} for the definition of $\mathcal{H},\mathcal{D}_{\hat{L}}$ for any operator $\hat{L}$.}
Here, the Hermitian $H_{\rm{tb}}  $ is a tight-binding Hamiltonian \red{inducing particle jumps between the two sites}. $\mathcal{L}_{\mathrm{D}}$ is responsible for dephasing at each site and   $\mathcal{L}_{\mathrm{E}}$ explicitly breaks the symmetry between the two sites and induces the occupation imbalance. Later on, it will be shown that controlling $\gamma$ allows to induce large $R$ values.  

The relation of eq.\eqref{eq:Lindblad dynamics} to the classical \red{asymmetric inclusion process} (ASIP) \red{\cite{Grosskinsky_ASIP}} is as follows. The dephasing and bias terms alone acting on the diagonal terms of the density matrix in the number basis lead to the ASIP master equation. Namly, the quantum master equation is split to the diagonal terms and the coherent terms, each having a closed set of equations (in the number operator basis).  The tight-binding Hamiltonian mixes between the cohrent terms and the diagonal terms, hence a quantum ASIP.

Three comments are in order before we present the results.  
First, note that here one cannot assume a priory that the quantum system is coupled to a series of thermalized baths as we have not performed a microscopic derivation of the Lindblad equation. Since we are not interested in studying thermalization, the lack of a microscopic derivation is of no importance.  Second, the steady state density matrix is not pure. Hence, we will only use the Logarithmic negativity as an entanglement quantifier. Third, in   eq.\eqref{eq:Lindblad dynamics} we have set $\hbar $ to unity. Furthermore, we will assume $\varepsilon,\eta,\gamma$ and the time $t$ to be dimensionless for convenience. When presenting the different scaling schemes, the inverse time dimensions of $\varepsilon,\eta,\gamma$ could be restored.

 The QASIP, like the Bose-Hubbard model can be shown to conserve the particle number $\hat{N} = \hat{n}_A + \hat{n}_B$ (see the appendix \ref{App:Lindblad operator dynamics}). However, a related but more general property exists for the QASIP.  In the number operator basis, we can write the density matrix as 
\begin{eqnarray}
\label{eq:spanning rhoS}
    \hat{\rho} &=&  \sum_{S} a_S \hat{\rho}_S, \quad \textrm{where} \\ \nonumber 
    \hat{\rho}_S &= &
     \sum^{S} _{x,y=0} \varrho_{S} (x,y) \ket{x,S-x} \bra{y,S-y}.
\end{eqnarray}
 Here, $S$ takes non-negative integer values and $a_S$ are non-negative prefactors that sum to $1$. Note that the hermitianity of the density matrix $\hat{\rho}_S$ implies $\varrho^* _S (x,y) = \varrho _S (y,x)$ and unity trace implies $\sum^S _{x=0} \varrho_S(x,x) =1$. 
 
 The dynamics of eq.\eqref{eq:Lindblad dynamics} \red{is restricted to} the subspace \red{of $\hat{\rho}_S$:} 
 \begin{eqnarray}
  \label{eq: S dynamics}
     \partial_t \varrho_S (x,y) &=& -\eta (x-y)^2  \varrho_S (x,y) \\ \nonumber &&
      +\gamma \,  x_- y_- \varrho_S (x-1,y-1)
     \\ \nonumber &&
     -\frac{1}{2}\gamma \left( x^2 _+ + y^2 _+ \right) 
     \varrho_S (x,y)
     \\ \nonumber &&
     -i \varepsilon  \sum_{z=\pm1 }
     x_z  \varrho_S (x+z,y)
     -y_z  \varrho_S (x,y+z)
     \\ \nonumber 
     X_+& = & \sqrt{(X+1)(S-X)} 
     \\ \nonumber 
     X_-& = & \sqrt{X(S-X+1)}, \quad \textrm{where } X=x,y. 
 \end{eqnarray}
 Namely, we have replaced the treatment of the infinite dimensional density matrix $\hat{\rho}$, with a treatment of finite dimensional $(S+1)^2$  density matrices $\hat{\rho}_S$ at fixed $S$.  
 
 The conserved number $S$ of $\hat{\rho}_S$ equals  the number of particles in the system as 
 \begin{equation}
 \Tr \hat{\rho}_S \hat{N}  = \sqrt{\Tr \hat{\rho}_S \hat{N}^2 }    = S .
 \end{equation}
 Eq.\eqref{eq: S dynamics} implies therefore that the process conserves the particle number. From hereon out, we may replace $S$ by $N$.

\begin{figure}
    \centering
    \includegraphics[scale=0.45]{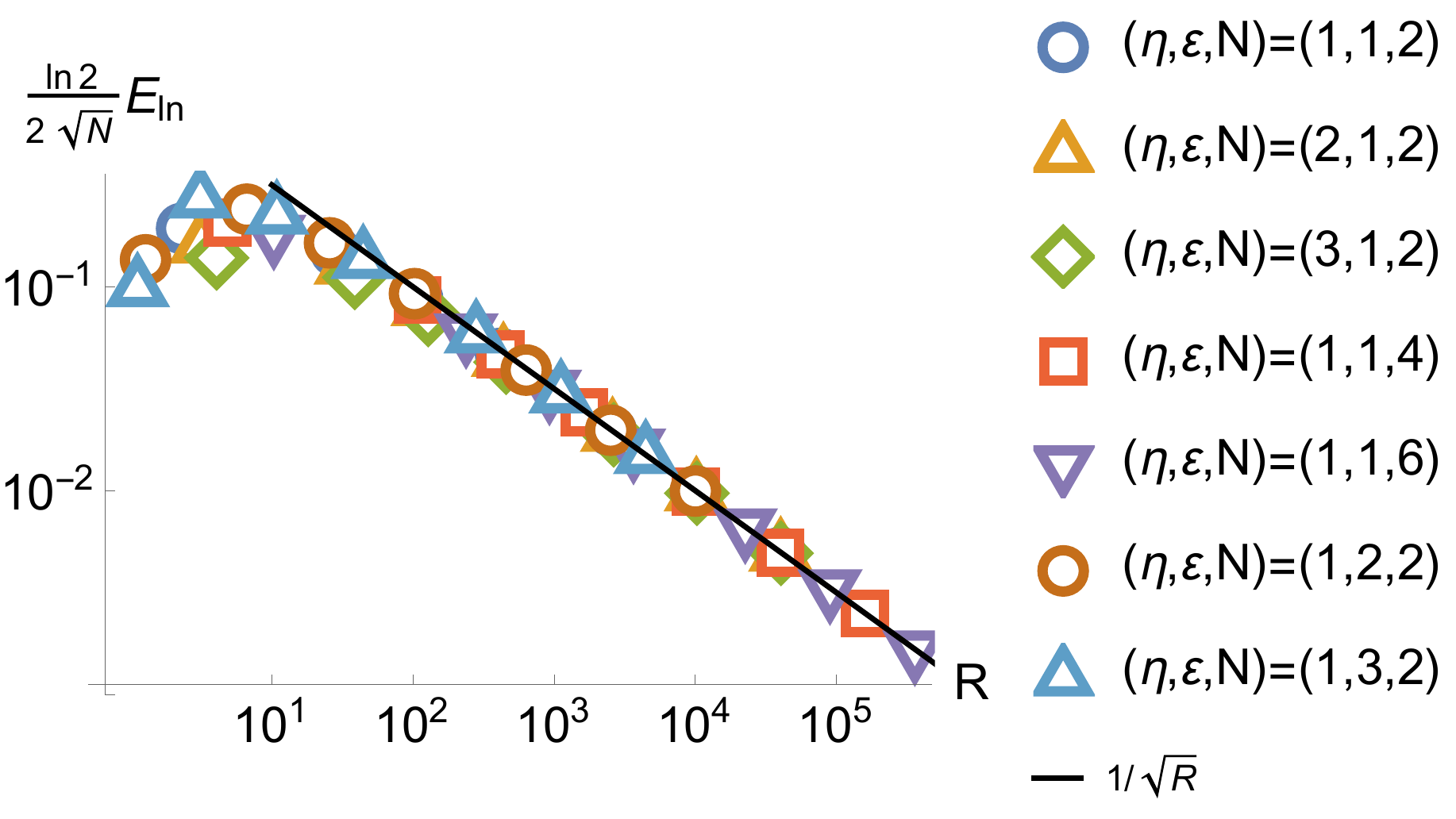}
    \caption{
    \red{Corroborating the perturbation theory analytical predictions presented in Table~\ref{table:QASIP model table}. The Logarithmic negativity is evaluated numerically for the QASIP model and compared with perturbation theory at the large $\gamma$ limit in the range $\gamma \in  \left[1,
    200\right]$. Different $\varepsilon,\eta,N$ values are considered, all showing the expected collapse onto the $1/\sqrt{R}$ plot at large $R$ values, with no fitting parameters.  }}
    %The scaled Logarithmic Negativity for the QASIP  with $\varepsilon=\eta=1, \gamma \in \left[1,200\right] $ and different $N$ values. The points collapse in the large $R$ limit onto a power law plot. The exponent is in accord with the analytical value of $\alpha =1/2$.   }
    \label{fig:QASIP large gamma different N }
\end{figure}

\begin{figure}
    \centering
    \includegraphics[scale=0.45]{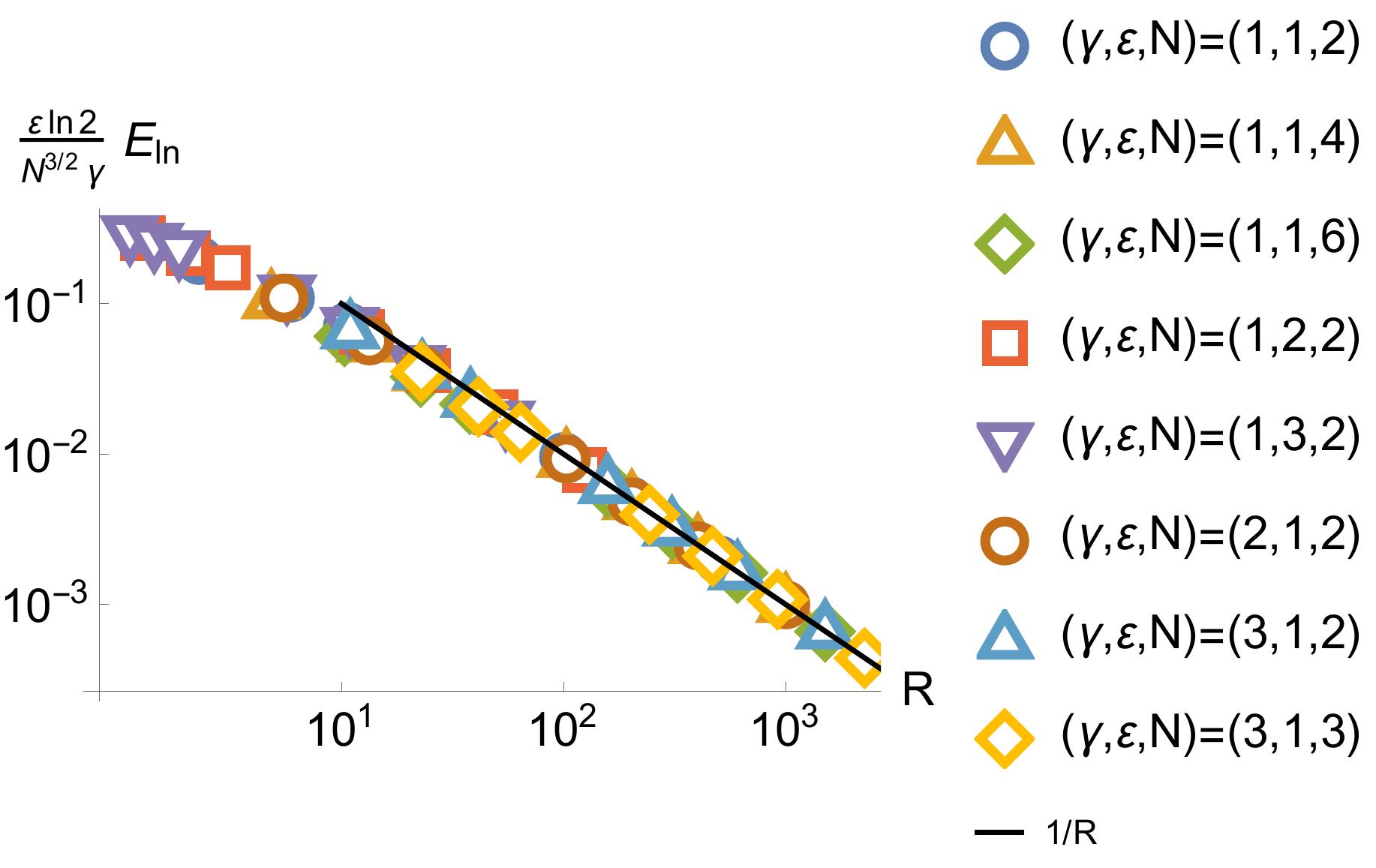}
    \caption{
    \red{Corroborating the perturbation theory analytical predictions presented in Table~\ref{table:QASIP model table}. The Logarithmic negativity is evaluated numerically for the QASIP model and compared with perturbation theory at the large $\eta$ limit in the range $\eta \in  \left[1,
    500\right]$. Different $\varepsilon,\gamma,N$ values are considered, all showing the expected collapse onto the $1/R$ plot at large $R$ values, with no fitting parameters.  }}
    %The scaled Logarithmic Negativity for the QASIP  with $\varepsilon=\gamma=1, \eta \in \left[1,500\right] $ and different $N$ values. The points collapse in the large $R$ limit onto a power law plot. The exponent is in accord with the analytical value of $\alpha =1$.   }
    \label{fig:QASIP large eta different N }
\end{figure}

\begin{figure}
    \centering
    \includegraphics[scale=0.45]{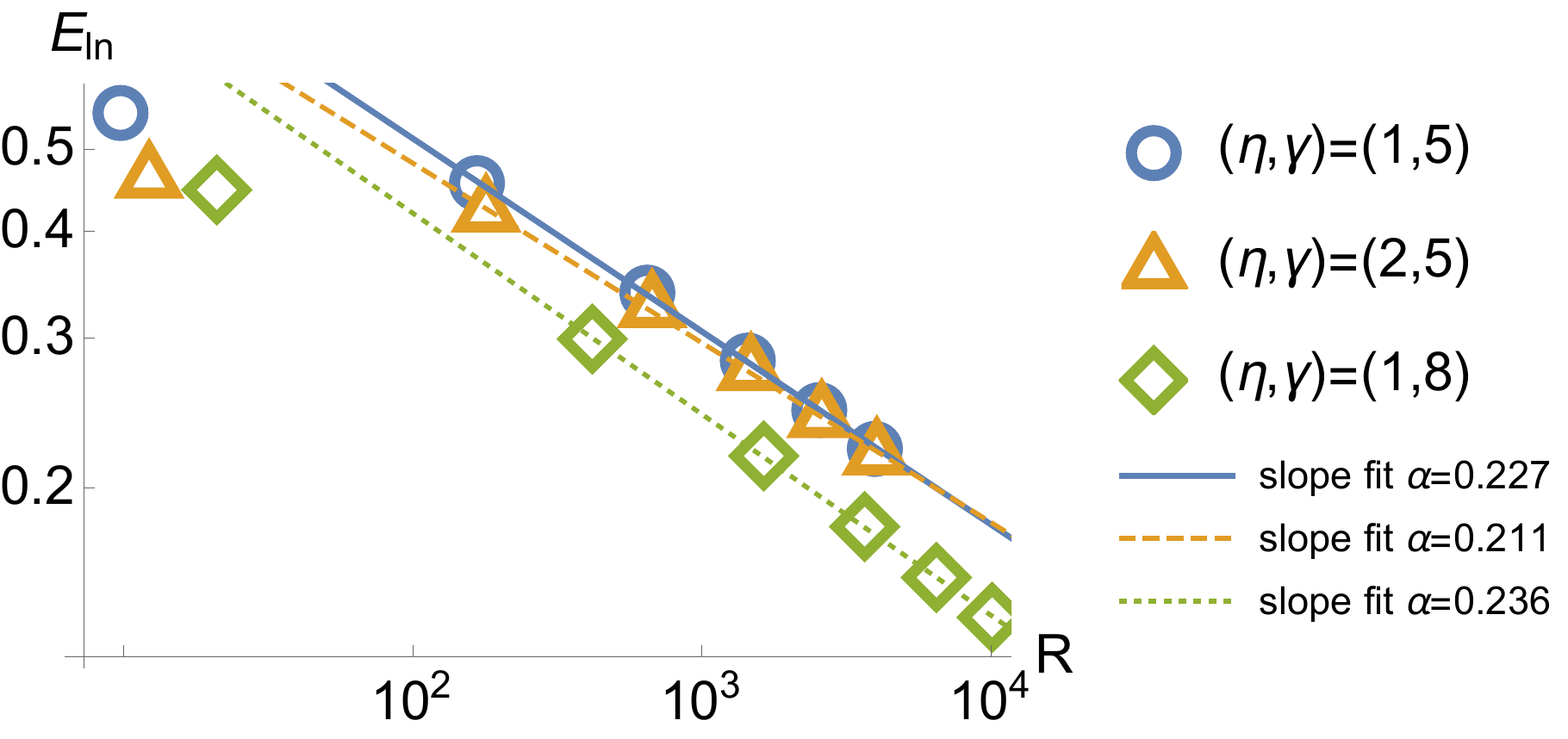}
    \caption{ 
    \red{Numerical fitting of the Logarithmic negativity at the large $N$ limit of the QASIP model. 
    Different $\eta,\gamma$ values were tested, 
leading to the large $R$ limit at $\varepsilon=1, N\in[1,25]$. Numerically, it is shown that $E_{ln} \propto R^{-\alpha}$, where 
$\alpha  \approx 0.236$.  }    }
    \label{fig:QASIP LogNeg plot }
\end{figure}

In eq.\eqref{eq:Lindblad dynamics}, there are different scaling schemes leading to the large $R$ limit \red{at the steady state}. The Table \ref{table:QASIP model table} summarises the power law behaviour of the Logarithmic negativity in three different scaling regimes. The results in Table \ref{table:QASIP model table}  were verified both analytically and numerically. See Fig.~\ref{fig:QASIP large gamma different N }, for the large $\gamma$ limit and Fig.~\ref{fig:QASIP large eta different N } for the large $\eta$ limit.  For the large $N$ only numerical evidence is currently present, see Fig.~\ref{fig:QASIP LogNeg plot } and \ref{fig:QASIP R of N scaling }.

\begin{table}[ht!]
\centering
\begin{tabular}{ |c||c|c|  }
 \hline
  &&\\ 
  & $R$ & $E_{ln}(R) = \Gamma R^{-\alpha} $ \\
  &&\\ 
 \hline
 &&\\ 
 $\gamma \gg 1$   & $\frac{\gamma ^2 N^2}{4 \varepsilon ^2}$    &$\Gamma = \frac{2 \sqrt{N}}{ \ln 2} , \quad \alpha =1/2$ \\
 && \\ 
 $\eta \gg 1$&   $\eta \frac{N \gamma }{2 \varepsilon^2}  $   & $\Gamma = \frac{2N^{3/2} \gamma }{\varepsilon \ln 2}, \quad \alpha =1 $   \\
 && \\ 
 $N \gg 1$ & $ \propto N^{\beta},\beta \approx 1.988 $ & $ \alpha \approx 0.236  $ \\
 \hline
\end{tabular}
\caption{The power law behaviour of the Logarithmic
 negativity in the steady state QASIP. 
 \red{For the large $N$ limit, the exponents are evaluated numerically.}  }
\label{table:QASIP model table}
\end{table}

In conclusion, the steady state QASIP in eq.\eqref{eq:Lindblad dynamics} exhibits a power law decay in the Logarithmic negativity, similarly to the Bose-Hubbard dynamics.

In the next section, we provide a detailed derivation of the results for the Bose-Hubbard model and for the QASIP.

%%% ============================================================================================================================================================================================ %%%

\section{Analytical and numerical analysis  \label{sec:Analytical treatment}}

In Sec.~\ref{sec:Results}, we have introduced two lattice models: the Bose-Hubbard model and the quantum asymmetric inclusion process. The power law behaviour of the  Von Neumann entanglement and the Logarithmic negativity was summarized in Tables \ref{table:BH model table} and \ref{table:QASIP model table}. In this section, we describe the analysis of these results in detail.

\subsection{Bose-Hubbard model}

To analyze the entanglement properties of the two-site Bose-Hubbard model in eq.\eqref{eq:BH Hamiltonian}, we need to find the $N$-particle ground state of the Hamiltonian. Given a description of the ground state, finding the ratio $R$ and the entanglement quantifiers $E_{vn}, E_{ln}$ becomes straight-forward, but sometimes technically cumbersome. See the appendix~\ref{App:Entanglement}.

The Hilbert space of $N$ particles for the Bose-Hubbard Hamiltonian eq.\eqref{eq:BH Hamiltonian} is spanned by  the $N+1$ Fock states $\ket{n_A,n_B} = \frac{1}{n_A ! n_B !}(\hat{b}^\dagger _A)^{n_A}(\hat{b}^\dagger _B)^{n_B}\ket{0,0}$. Namely, 
\begin{equation}
    \ket{\psi_N} = \sum^N _{k=0} a_k \ket{k,N-k}, \quad \textrm{where }\sum^N _{k=0}|a_k|^2 =1.
\end{equation}
Then, the $N$ particle Bose-Hubbard Hamiltonian can be written as a $N+1 \times N+1 $ matrix. Finding the ground state can be done analytically and for all $\mu,U,J$ values when we set $N=1$. This simple case will reveal the intuitive scaling limits leading to the large $R$ behaviour, both for the Bose-Hubbard model as well as for the quantum asymmetric inclusion process in the next subsection. 

Indeed, for $N=1$, the Bose-Hubbard Hamiltonian can be represented by the $2\times 2$ matrix 
\begin{equation}
\label{eq:BH Hamiltonian for N=1}
    H^{(N=1)} _{BH} = \begin{pmatrix}
-\mu  & -J \\
-J & 0 ,
\end{pmatrix}
\end{equation}
where the wavefunction is in its most general form
\begin{equation}
\label{eq:ket state for BH with N=1 }
\ket{\psi_{N=1}} = \cos \zeta \ket{10} + {\rm e}^{i \phi_1 } \sin \zeta \ket{01}   = \begin{pmatrix}
\cos \zeta \\
{\rm e}^{i \phi_1} \sin \zeta 
\end{pmatrix}   
\end{equation}
 for real $\zeta, \phi_1 $ values and the right hand side of eq.\eqref{eq:ket state for BH with N=1 } is in a vector notation corresponding to the matrix in eq.\eqref{eq:BH Hamiltonian for N=1}.  Clearly, the ratio is $R = \cot^2 \zeta$. The lowest eigenvalue of eq.\eqref{eq:BH Hamiltonian for N=1} is $\epsilon = - \frac{1}{2} (\mu + \sqrt{\mu^2 + 4J^2} )$ with the ground state $\ket{\psi_{N=1}} = \frac{1}{\mathcal{N}} ( -\epsilon \ket{10} +J \ket{01} )$ and \red{$\mathcal{N}^2 = \epsilon^2 +J^2$} is a normalization constant. This implies that $R = \epsilon^2/J^2  $. Therefore, in this particular case, $R \gg 1$ only if $\mu/J \gg 1$ and leads to $R = (\mu/J)^2 + O(\mu)  $ asymptotically.  

The Von Neumann entanglement for the ground state in eq.\eqref{eq:ket state for BH with N=1 } is 
\begin{equation}
    E_{vn} = - 2 \cos^2 \zeta \ln (\cos \zeta) - 2 \sin^2 \zeta  \ln (\sin \zeta)  . \end{equation}
Using trigonometric identities, we recover \red{$\cos^2 \zeta = \frac{R}{1+R}$ and $\sin^2 \zeta = \frac{1}{1+R}$.} Asymptotically for large $R$,  \red{$E_{vn} =  \frac{\ln R}{R} $ to leading order} as reported in Table \ref{table:BH model table}. 

To find the Logarithmic negativity, we need to write the partially transposed density matrix 
\red{
\begin{eqnarray} 
    \rho^{PT} &=& \cos^2 \zeta \ket{10}\bra{10} + \sin^2 \zeta \ket{01}\bra{01}  \\ \nonumber 
    && + \cos \zeta \sin \zeta \left( {\rm e}^{i\phi_1} \ket{00}\bra{11}  + {\rm e}^{-i\phi_1} \ket{11}\bra{00}  
    \right).
\end{eqnarray}
}
The eigenvalues of the partially transposed density matrix are $\cos^2 \zeta , \sin^2 \zeta$ and $\pm  \frac{\sqrt{1-\cos 4\zeta }}{2\sqrt{2}}$. Therefore, the Logarithmic negativity is \red{$E_{ln} = \log_2 (1 +\frac{\sqrt{1-\cos 4\zeta }}{\sqrt{2}})  $.} For large $R$ and to leading order, we find $E_{ln} = \frac{2}{\ln 2} R^{-1/2}  + O(R^{-1})  $  as reported in Table~\ref{table:BH model table}.

For $N>1$, the ground state solution becomes cumbersome, but still requires dealing with a  $N+1\times N+1   $ Bose-Hubbard matrix. The numerical code that produced Fig.~\ref{fig:BH large mu},\ref{fig:BH large N fitting} and \ref{fig:BH RofN} finds the ground state of the Bose-Hubbard matrix at some finite $N$. Then, it calculates the Von Neumann entanglement and the Logarithmic negativity.  

From the $N=1$ example , we have seen that the large $\mu$ limit leads to a large bias $R \gg 1$. This happens when the $\mu$ term  dominates the energy of the ground state, i.e. for  $\mu \gg \sqrt{N}J , U N^2$. Another large $R$ limit is recovered for finite $\mu,U,J > 0$ and large $N$. In this limit, 
the particles condense due to the strong attractive energy $ \sim U N^2  $. The symmetry is broken by the potential offset $\mu$, leading to condensation of the particles in site $A$ and to large $R$ values.

A perturbative approach for the large $N$ limit is non-trivial due to \red{as the dimension of the effective Hilbert space changes with $N$}. However, a direct numerical analysis clearly reveals the power law behaviour in this limit. See Table \ref{table:BH model table}, Sec.\ref{sec:Results} and the appendix \ref{app:BH model} for more details. In what follows, we consider the large $\mu$ limit and evaluate the ground state and the entanglement quantifiers using a perturbative approach.

% The $N$-particle eigenstates of  eq.\eqref{eq:BH Hamiltonian} are spanned by the $N+1$ Fock states $\ket{n_A,n_B} = \frac{1}{n_A ! n_B !}(\hat{b}_A)^{n_A}(\hat{b}_B)^{n_B}\ket{0,0}$. Namely, 
% \begin{equation}
%     \ket{\psi_N} = \sum^N _{k=0} a_k \ket{k,N-k}, \quad \textrm{where }\sum^N _{k=0}|a_k|^2 =1.
% \end{equation}
% The eigenstates satisfy $H_{BH} \ket{\psi_N} = \epsilon \ket{\psi_N} $. To obtain the entanglement quantifiers $E_{vn},E_{ln}$ as a function of $R$, we need to find the $N$ particle ground state $\ket{\Psi_N}$. Namely, the lowest value of $\varepsilon $ such that $H_{BH}\ket{\Psi_N}= \varepsilon\ket{\Psi_N}$. 

Let us develop a standard perturbation theory for the Hamiltonian $H = H_0 + \frac{1}{\mu} H_1 \red{= \frac{1}{\mu}H_{BH}} $
\begin{equation}
    H_0 = - \hat{n}_A   \quad H_1  = H_{BH} -\mu  H_0 , 
\end{equation}
at large $\mu $.

The eigenstates of $H_0$ are $\ket{\phi^{(0)} _n} = \ket{n,N-n}$ with energies $\epsilon^{(0)} _n = - n$. The first order correction to the ground state is  
\begin{equation}
    \ket{\phi_N} = \ket{N,0} + \lambda \ket{N-1,1} + O(\lambda^2)
\end{equation}
where $\lambda = \frac{J \sqrt{N}}{\mu}$ assumed to be small as well as $U N^2 / \mu   \ll 1 $. At this limit we find $R = \langle \hat{n}_A \rangle  / \langle \hat{n}_B \rangle  = N / \lambda^2.  $ So, we can approximate at small $\lambda$, $R = \mu^2 /J^2 + O(\mu)$. The Von Neumann entanglement can be calculated for the ground state $\ket{\phi_N}$
\begin{eqnarray}
    E_{vn} = - \frac{1}{1+\lambda^2} \ln \frac{1}{1+\lambda^2} - \frac{\lambda^2}{1+\lambda^2} \ln \frac{\lambda^2}{1+\lambda^2} 
    \\ \nonumber 
    = - \lambda^2 \ln \lambda^2   + O(\lambda^3).
\end{eqnarray}
At this limit, we find the reported scaling 
\begin{equation}
\label{eq:Evn large mu}
E_{vn} = \frac{N}{R} \ln \frac{N}{R} 
\end{equation}
to leading order. This is corroborated numerically in Fig.~\ref{fig:BH large mu} and summarised in Table \ref{table:BH model table}. Recall that in this scaling, $N$ may be large, but $R \gg N$. 

To find the Logarithmic negativity, we need to calculate the eigenvalues of the partially transposed density matrix of $\ket{\phi_N}$. For $N>1$, the only non-zero eigenvalues are $\frac{1}{1+\lambda^2},\frac{\lambda^2}{1+\lambda^2},\frac{\pm \lambda }{1+\lambda^2}$. This leads to 
\begin{equation}
\label{eq:Eln large mu}
    E_{ln} = \log_2 \left( 1+ \frac{2\lambda }{1+\lambda^2} \right) = \frac{2}{\ln 2} \sqrt{\frac{N}{R}} + O(1/R). 
\end{equation}
    For large $R$ values, where the perturbation theory applies, the Logarithmic negativity dominates the Von Neumann entanglement as it should \cite{Plenio_Entanglement_Review,plenio2005entropy}. Again, we refer to Fig.~\ref{fig:BH large mu} to see the excellent agreement with the numerical evaluation.

Other scaling schemes, leading to large $R$ values can exist. Nevertheless, the power law behaviour of the entanglement quantifiers is believed to persist, based on the $N=1$ exactly solvable cases. 

We turn to study the large $R$ entanglement properties of a completely different setup -- the quantum asymmetric inclusion process.

\subsection{The QASIP}

% \begin{itemize}
%     \item Describe the state space of the density matrix at fixed S. 
%     \item routine: find the steady state at fixed S. Calculate the PT density matrix and then the LogNeg. 
%     \item example: S=1. Exact analytic treatment. Identify scaling schemes. 
%     \item The large $\gamma$ limit 
%     \item The large $\eta$ limit. \item The large $S$ limit. 
% \end{itemize}

To analyze the \red{steady state} entanglement properties of the QASIP at large $R$, we need to find the steady state density matrix with a fixed $S$, i.e.  $\mathcal{L}_{QASIP}(\hat{\rho}_S) =0 $. Namely, we wish to find $\varrho_S (x,y)$ such that the right hand side of eq.\eqref{eq: S dynamics} vanishes.

%  \underline{$S=0$:} The simplest case leads to $x,y=0$ which gives the single  equation 
%  \begin{equation}
%      \partial_t \varrho_0 (0,0) =  0. 
%  \end{equation}
%  Namely, we find the time-independent solution $\hat{rho}_0 = \ket{0,0}\bra{0,0}$. 

As in the Bose-Hubbard model, it is useful to study first the simple case of $S=1$.  Here, $x,y=\lbrace 0,1 \rbrace$ and demanding a steady state in eq.\eqref{eq: S dynamics} leads to
\begin{equation}
\label{eq:QASIP S=1 matrix}
    \begin{pmatrix}
-\gamma & i \varepsilon & -i \varepsilon & 0\\
i \varepsilon & -\frac{1}{2}\gamma-\eta  & 0 & -i \varepsilon\\
-i \varepsilon & 0 & -\frac{1}{2}\gamma -\eta  & i \varepsilon\\
\gamma & -i \varepsilon & i \varepsilon & 0
\end{pmatrix} 
\begin{pmatrix}
\varrho_1 (0,0) \\
\varrho_1 (0,1) \\
\varrho_1 (1,0) \\
\varrho_1 (1,1) 
\end{pmatrix} = 
\begin{pmatrix}
0 \\
0 \\
0 \\
0 
\end{pmatrix}.
\end{equation}
Solving eq.\eqref{eq:QASIP S=1 matrix}, we find the steady state solution for $S=1$ 
\begin{eqnarray}
     \mathcal{N}_1  \hat{\rho}_1  &=&   
     4 \varepsilon^2  \ket{0,1}\bra{0,1} 
     \\ \nonumber && 
     + 2i\gamma \varepsilon \left(\ket{1,0}\bra{0,1}-\ket{0,1}\bra{1,0}\right)  
     \\ \nonumber && 
     + (4\varepsilon^2 + \gamma^2 +2 \gamma \eta) \ket{1,0}\bra{1,0}.
\end{eqnarray}
where $\mathcal{N}_1 = \gamma ^2+2 \gamma  \eta +8 \varepsilon ^2 $ ensures unity trace of $\hat{\rho}_1$. From the steady state solution we recover $R = 1 + \frac{\gamma(\gamma+2\eta)}{4\varepsilon^2} $ and $E_{ln} = \log_2 \left( 1+ \frac{4\gamma \varepsilon}{\gamma^2 + 8 \varepsilon^2 + 2\gamma \eta}
\right) $. A few observations can already be made. At $\eta \rightarrow \infty$ the entanglement vanishes as can be expected in the large dephasing limit \cite{Bernard18,Altman_MBL}. Moreover, the Logarithmic negativity becomes positive due to a combination of biasing and coherent hopping, i.e. $\gamma \varepsilon>0$. We identify two limits where $R$ becomes large.  For $\gamma \rightarrow \infty $ and finite $\eta,\varepsilon$ we recover $R \propto \gamma^2 $ and $E_{ln} \propto 1/ \sqrt{R} $. Similarly, for finite \red{$\gamma,\varepsilon$} and large $\eta$ we recover $R \propto \eta$ and $E_{ln}\propto 1/ R$. Already for the $S=1$ case we find that the entanglement power law behavior persists for large $R$ values. However, the exponent is non-universal and depends on the scaling scheme.

Exact solution of the steady state for $S>1$ is at best tedious. Instead, we will find the steady state solution at the two limits noted above, using a perturbative approach; The limits at finite $S$ will be shown to agree with the $S=1$ exact analysis carried out in the above. 

% ================================================================================ %
\subsubsection{Large asymmetry between the sites}

For $\gamma \gg 1$, and at finite $S,\eta,\varepsilon$, we develop the steady state density matrix as a perturbative sum 
\begin{equation}
\label{eq:gamma pert dm}
    \hat{\rho}_S = \frac{1}{\mathcal{N}_{\gamma}}(\hat{\rho}^{(0)} _S + \frac{1}{\gamma} \hat{\rho}^{(1)} _S + \frac{1}{\gamma^2} \hat{\rho}^{(2)} _S )  + O(1/\gamma^3),
\end{equation}
where $\mathcal{N}_{\gamma}$ is a normalization constant to assure trace one of the truncated density matrix. This perturbative approach implies the order by order steady state solutions
\begin{eqnarray}
 \label{eq:gamma order 0}
 0 &=& \mathcal{L}_{\mathrm{E}}  (\hat{\rho}^{(0)} _S) 
  \\  \label{eq:gamma order 1}
 0 &=& \frac{1}{\gamma}\mathcal{L}_{\mathrm{E}} (\hat{\rho}^{(1)} _S) + (\mathcal{L}_{\mathrm{D}} + \mathcal{H}_{tb}) \hat{\rho}^{(0)}_S   
  \\ \label{eq:gamma order 2}
 0 &=& \frac{1}{\gamma}\mathcal{L}_{\mathrm{E}} (\hat{\rho}^{(2)} _S) + (\mathcal{L}_{\mathrm{D}} + \mathcal{H}_{tb}) \hat{\rho}^{(1)}_S   
\end{eqnarray}
Eq.\eqref{eq:gamma order 0} admits a uniqe solution   $\hat{\rho}^{(0)} _S = \ket{S,0}\bra{S,0} $, namely to leading order site $A$ is maximally occupied and site $B$ is depleted. Using the leading order solution, we find $\hat{\rho}^{(1)} _S = \frac{2 i \varepsilon}{\sqrt{S}} (\ket{S,0}\bra{S-1,1} + \ket{S-1,1}\bra{S,0})$. Note that to first order, there are yet no corrections to the occupancies. Hence, we solve to second order in $1/\gamma $ obtaining 
\begin{eqnarray}
\hat{\rho}^{(2)} _S &=& \frac{4 \varepsilon ^2}{\gamma ^2 S} \ket{S-1,1}\bra{S-1,1}      
 \\ \nonumber 
 &+&  \frac{4i \varepsilon  \eta }{S^{3/2}} \ket{S-1,1}\bra{S,0} %- \ket{S,0}\bra{S-1,1})
 \\ \nonumber 
  &-& \frac{2 \sqrt{2} \varepsilon ^2}{\sqrt{S (S-1)}} \ket{S,0}\bra{S-2,2} 
  %+ \ket{S-2,2}\bra{S,0})
  + \textrm{h.c.}
  .
\end{eqnarray}
To second order, we find $\mathcal{N}_{\gamma} = \frac{4 \varepsilon ^2}{\gamma ^2 S}+1$. From the perturbative solution  of eq.\eqref{eq:gamma pert dm}, we find that
\begin{equation}
R =  S-1 + \frac{S^2 \gamma^2}{4\varepsilon^2} + O(\gamma) \approx  \frac{S^2 \gamma^2 }{4\varepsilon^2} .
\end{equation}
This approximation also implies the assumption $\varepsilon^2 \ll S \gamma^2$. Also, the Logarithmic negativity can be calculated as there are at most four non-zero eigenvalue for partially transposed density matrix for any $S$ value. We find  to leading order \begin{equation}
    E_{ln} = \frac{4  \varepsilon }{\gamma  \sqrt{S} \ln 2}
    +O(\frac{1}{\gamma ^2})
    \approx \frac{2 \sqrt{S}}{ \ln 2 }\frac{1}{\sqrt{R}} + O(\frac{1}{R}).
\end{equation}
As noted in Sec.~\ref{sec:Results}, the $E_{ln}(R)$ power law behaviour was verified numerically \red{in Fig.~\ref{fig:QASIP large gamma different N }}.

% ================================================================================ %
\subsubsection{Large dephasing limit}

Here we consider the large $\eta$ limit with fixed $S,\gamma, \varepsilon$. We write the density matrix as a perturbative series in $1/\eta$ 
\begin{equation}
    \label{eq:eta pert dm}
    \hat{\rho}_S = \frac{1}{\mathcal{N}_{\eta}}(\hat{\rho}^{(0)} _S + \frac{1}{\eta} \hat{\rho}^{(1)} _S + \frac{1}{\eta^2} \hat{\rho}^{(2)} _S )  + O(1/\eta^3),
\end{equation}
where here $\mathcal{N}_\eta$ is a normalization constant ensuring the truncated density matrix has trace $1$.   Again, the perturbative series  implies the order by order steady state solutions
\begin{align}
 \label{eq:eta order 0}
 0 =& \mathcal{L}_{\mathrm{D}}  (\hat{\rho}^{(0)} _S) 
  \\  \label{eq:eta order 1}
 0 =& \frac{1}{\eta}\mathcal{L}_{\mathrm{D}} (\hat{\rho}^{(1)} _S) + (\mathcal{L}_{\mathrm{E}} + \mathcal{H}_{tb}) \hat{\rho}^{(0)}_S   
  \\ \label{eq:eta order 2}
 0 =& \frac{1}{\eta}\mathcal{L}_{\mathrm{D}} (\hat{\rho}^{(2)} _S) + (\mathcal{L}_{\mathrm{E}} + \mathcal{H}_{tb}) \hat{\rho}^{(1)}_S   
\end{align}
Eq.\eqref{eq:eta order 0} admits a degenerate solution
\begin{equation}
\hat{\rho}^{(0)} _S  = \sum^S _{k=0} a_k \ket{k,S-k}\bra{k,S-k}  
\end{equation}
 with $a_k $ non-negative coefficients. This degeneracy is broken in the next order, i.e.  eq.\eqref{eq:eta order 1}. We find  $\hat{\rho}^{(0)} _S  = \ket{S,0}\bra{S,0}$, however the degeneracy moves to the next order
\begin{align}
\hat{\rho}^{(1)} _S  =&
\sum^S _{k=0} b_k \ket{k,S-k}\bra{k,S-k}
 \\ \nonumber 
+&i\varepsilon \sqrt{S} (\ket{S,0}\bra{S-1,1}-\ket{S-1,1}\bra{S,0}) , 
\end{align}
where $b_k$ are again non-negative coefficients. To evaluate to leading order $R$, we have to break the degeneracy in $b_k$. This breaking is obtained at the next order, i.e. eq.\eqref{eq:eta order 2}, where we find $b_k = \delta_{k,S-1} \frac{2\varepsilon^2}{\gamma}$ and      
\begin{align}
\hat{\rho}^{(2)} _S  =&
-\frac{\varepsilon ^2 \sqrt{S-1} \sqrt{S}}{2 \sqrt{2}} \ket{S-2,2}\bra{S,0}
 \nonumber  \\ \nonumber 
+&i\frac{\gamma ^2 \varepsilon  S^{3/2}+4 \varepsilon ^3 \sqrt{S}}{2 \gamma } \ket{S-1,1}\bra{S,0}
 \\ \nonumber 
 -&i\frac{2 \sqrt{2} \varepsilon ^3 \sqrt{S-1}}{\gamma } \ket{S-2,2}\bra{S-1,1}
 \\ \nonumber 
 +& \sum^S _{k=0} c_k \ket{k,S-k}\bra{k,S-k}
 \\ 
& +\textrm{h.c.}.
\end{align}
The degeneracy in the non-negative terms $c_k$ is broken at the third order of the expansion. To leading order in $\eta$, we find $\mathcal{N}_\eta = 1+\frac{2 \varepsilon ^2}{\gamma  \eta } $. Therefore, to leading order    
\begin{equation}
    R=\frac{\gamma  \eta  S}{2 \varepsilon ^2}+S-1\approx \frac{\gamma  \eta  S}{2 \varepsilon ^2}.
\end{equation}
Again, the spectrum of the partially transposed density matrix is composed of only four non-zero eigenvalues for any $S>1$: $(-\frac{ \gamma  \varepsilon  \sqrt{S}}{\gamma  \eta +2 \varepsilon ^2},\frac{ \gamma  \varepsilon  \sqrt{S}}{\gamma  \eta +2 \varepsilon ^2},\frac{2 \varepsilon ^2}{\gamma  \eta +2 \varepsilon ^2},\frac{\gamma  \eta }{\gamma  \eta +2 \varepsilon ^2})$. The Logarithmic negativity is thus given by     
\begin{equation}
    E_{\ln }=\log_2\left(1+\frac{2 \gamma  \varepsilon  \sqrt{S}}{\gamma  \eta +2 \varepsilon ^2}\right)\approx \frac{2 \varepsilon  \sqrt{S}}{\eta  \ln 2 }=\frac{ \gamma  S^{3/2}}{ \varepsilon  \ln 2 } \frac{1}{R}.
\end{equation}

\red{As noted in Sec.~\ref{sec:Results}, the $E_{ln}(R)$ power law behaviour was verified numerically in Fig.~\ref{fig:QASIP large eta different N }}.

\subsection{Large number of particles}

% \underline{$S=2$:}
% The exact expressions for the occupancy ratio $R(\gamma,\eta,\varepsilon)$ and the Logarithmic negativity $E_{ln} (\gamma,\eta,\varepsilon)$ is cumbersome and is given in the \ref{app:sec:S2 expression}. However, the two previous limits give the same scaling behaviour. Namely, 
% for $\gamma \gg 1$ and $\eta,\varepsilon$ finite, we find $R\propto \gamma^2 \gg 1$ and $E_{ln}\propto 1/ \sqrt{R}$. For $\eta \gg 1 $ and $\varepsilon,\gamma$ finite, we find as before $R \propto \eta \gg 1 $ and $E_{ln} \propto 1/R$. 

%\underline{$S>2$:}
% Analytical expressions can still be obtained for rather large $S$ values. However, the expressions are cumbersome already at $S=2$ \blue{as can be seen in appendix \ref{app:sec:S2 expressions}}. At the large $\gamma$ limit, we observe that $R = \frac{S^2 \gamma^2}{4 \varepsilon^2 }  +O(\gamma) $ and $E_{ln} = \frac{4}{\sqrt{S} \log 2} \frac{\varepsilon}{\gamma} +O(1/\gamma)$. This implies that for $\gamma \gg 1 $, we have $E_{ln} = \frac{2 \sqrt{S}}{\log 2} \frac{1}{R^{1/2}} $ which again demonstrates the power law behaviour of $E_{ln}(R)$. The second limiting case, for large $\eta$, we find $R = \frac{S \gamma \eta^2}{2 \varepsilon^2} +O(1)$ and $E_{ln} = \frac{2\sqrt{2}}{\log 2}\frac{\varepsilon}{\eta}+O(1/\eta)$. This implies $E_{ln} = \frac{\gamma S^{3/2}}{\varepsilon \log 2 } \frac{1}{R} $. These limiting cases where calculated analytically for $S=1,2,3,4$. 

We also studied the scaling limit $S \gg 1$ and finite $\eta,\gamma,\varepsilon$. Analytically, a perturbative solution in this case becomes hard due to \red{the change in the state space, similarly to the
 Bose-Hubbard case}. Nevertheless, it is possible to numerically find the steady state and calculate the Logarithmic negativity even for large $S$ values. This was carried out numerically \red{(Fig.~\ref{fig:QASIP LogNeg plot })} and reported on in Sec.~\ref{sec:Results}.

\section{Discussion \label{sec:Discussion} }

State of the art experimental techniques allows to entangle a single agent to  thousands of atoms \cite{McConnell2015}. However, it was unclear whether one could push the experimental techniques to significantly increase the number of atoms entangled to the agent. 

Here, we have explored the theoretical bounds on entangling one or a few agents to a many body system. The ground state of a two-site Bose-Hubbard model, with an occupancy bias $R\gg1 $ leads to a power law decay in the Logarithmic negativity and the Von Neumann entanglement entropy in different scaling limits. Furthermore, the steady state of the QASIP biased to large $R$ values also exhibits a power law decay in the Logarithmic negativity. We stress that while the power law behaviour is typical,  the exponent depends on the scaling limits, see Tables \ref{table:BH model table} and \ref{table:QASIP model table}.

From the slow decay of the entanglement, it is now clear  it is typically possible to entangle thousands of atoms to a single agent. %Finding the exact conditions required to attain the observed  power law behaviour   is an open challenge.
Furthermore, designing systems with slow entanglement decay (small $\alpha$) allows to entangle more particles in the many body system to the one agent (or a few).   \red{It would be particularly appealing to develop a perturbative approach to the large $N$ limit in both models. Such an approach would allow to extract the exponents analytically, explore their range and dependence on the model parameters. Furthermore, it would suggest how best to tune the parameters to entangle the diluted system $B$ to the highly occupied system $A$.  }

The average Von Neumann entanglement entropy over the random pure state of Hilbert space $N \times N$ is $E_{vn} \sim \log N$ \cite{Page_averageEnropy,sen1996average}. Therefore, the Von Neumann entanglement entropy of the ground state is fundamentally different than that of the average. This is not too surprising when one relates to the area law of ground states in extended systems compared to the typical volume law. In turn, the low Von Neumann entanglement entropy of the ground state suggests that ground states in the large $R$ limit could be susceptible to analytical and numerical techniques, even for large many body systems. 

From the analysis so far, it may seem that the two-site lattice model is paramount to achieve the power law behaviour. We have carried out preliminary tests in a three site Hubbard model. Taking sites $A,B$ to occupy most of the particles in the system, namely $R = \langle \hat{n}_A + \hat{n}_B \rangle / \langle \hat{n}_C \rangle  $. The Von Neumann entanglement entropy between site $C$ and the subsystem $AB$  still exhibits a power law in large $R$. The analysis is beyond the scope of this work and will be presented elsewhere. 

Another question that comes to mind is whether the power law behaviour persists also in continuum models, and not only in lattice models. We believe this is not the case.  After coarse graining a lattice model into a continuum model, an increase is expected in the Von Neumann entanglement entropy due to loss of information. This increase does not depend on the occupancies and hence adds a constant to the Von Neumann entanglement entropy.  Therefore, in the large $R$ limit we expect to observe a saturation to a constant with a power law correction. Naively, that should be the same power law of the lattice model. It would be interesting to test this conjecture in future works.

Acknowledgments: I would like to thank Guy Cohen, Shahaf Asban and Ofir E. Alon for stimulating talks on the subject.

 % ==================================================================================================================================================================================== % 

 \appendix 
 
 % ==================================================================================================================================================================================== % 

\section{Entanglement quantifiers \label{App:Entanglement}}

In this section, we provide a brief introduction to the entanglement quantifiers used in this text: the Von Neumann entanglement entropy and the Logarithmic negativity.  The purpose of quantifiers is to distinguish between entangled to non-entangled states (separable) and furthermore to suggest a hierarchy of values for entangled states. Here we do not aim to give an exhaustive account of quantum quantifiers, but to motivate the usage of the Von Neumann entanglement entropy and the Logarithmic negativity in the case at hand.

For pure states, all entanglement measures are defined to correspond to the Von Neumann entanglement entropy \cite{Plenio_Entanglement_Review}. In bipartite system $AB$, 
\begin{equation}
    E_{vn}(\rho_{AB}  ) = -\Tr \rho_A \ln \rho_A  =  -\Tr \rho_B \ln \rho_B ,
\end{equation}
where $\rho_A = \Tr_B \rho_{AB}$ is the reduced density matrix. $E_{vn}>0$ only for non-separable pure states. 

In terms of wave functions (which are pure states), the Schmidt decomposition using orthonormal states imply $\ket{\psi} = \sum_i \alpha_i \ket{u_i}_A \otimes \ket{v_i}_B$. Then, we find $E_{vn}(\ket{\psi}) = - \sum_i |\alpha_i|^2 \log |\alpha_i|^2  $.

Entanglement is harder to quantify for mixed states. Many different measures for the entanglement exists. Typically, entanglement measures are given in the form of some minimization problem, making them hard to calculate. Instead, we will use the Logarithmic negativity which is an entanglement monotone and not a measure. Namely, for pure state the Logarithmic negativity does not correspond to the Von Neumann entanglement entropy (except for specific cases). However, it is straight-forward to calculate the  Logarithmic negativity, making it a favorable entanglement quantifier.

The Logarithmic Negativity is given by 
\begin{equation}
    \label{eq:LogNeg}
    E_{ln} (\rho) =   \log_2 \lVert \rho^{PT} \rVert_1,
\end{equation}
where $ \rho^{PT}$ is the partially transposed density matrix, and $\lVert A \rVert_1 \equiv  \Tr \sqrt{A A^\dagger} $. Intuitively speaking, the Logarithmic negativity counts the amount of negative eigenvalues in the partially transposed density matrix relating it to the Peres–Horodecki criterion \cite{Peres1996,HORODECKI1996}. We note that a positive Logarithmic negativity values insures non-separability, but a vanishing value does not guarantees separability.

The Logarithmic Negativity is an entanglement monotone  \cite{Plenio_LogNeg,Plenio_Entanglement_Review,Huber_EntanglementCost_Review}, which implies that on average, under locally quantum operations and classical communication (LOCC), the Logarithmic Negativity does not increase.  Furthermore, the Logarithmic negativity was shown to be an upper bound for the distillation entanglement, connecting it to useful quantum operations using maximally entangled states \cite{EntanglementCost_LogNeg}. Since the distillation entanglement is an entanglement measure, it is evident that for pure states, the Von Neumann entanglement entropy is bounded by the Logarithmic negativity. This fact provides a consistency check in our numerical assessment.

% Note that Logarithmic negativity is not an entanglement measure, as it generally does not match the Von Neumann entropy of entanglement for pure states. A spacial case where the Logarithmic negativity does match the Von Neumann entanglement entropy is for maximally entangled states.

% =============================================================================================================================================================================== % 

 \section{The Lindblad adjoint dynamics \label{App:Lindblad operator dynamics}}

The purpose of this section is to introduce the Heisenberg operator evolution picture for the Lindblad dynamics. 

For an observable $\hat{O}$ (explicitly time-independent), we have the expectation value $\langle \hat{O} \rangle  = \Tr \hat{O} \rho $. Therefore, 
\begin{equation}
    \partial_t \langle \hat{O} \rangle  = \Tr \hat{O} \partial_t\rho =  \Tr \hat{O} \mathcal{L}(\rho), 
\end{equation}
where $\mathcal{L}(\rho)$ is a Lindblad super-operator of  \red{eq.}\eqref{eq:GKSL equation general}. Then, the formal adjoint $\mathcal{L}^\dagger$ is defined such that 
\begin{equation}
    \partial_t \langle \hat{O}\rangle  = \Tr \mathcal{L}^\dagger(\hat{O}) \rho.  
\end{equation}
 For the Lindblad super-operator in eq.\eqref{eq:GKSL equation general}, it implies the Heisenberg picture 
 \begin{equation}
     \partial_t \hat{O} = \mathcal{L}^\dagger (\hat{O} )= - \mathcal{H}(\hat{O}) + \sum_k \hat{L}^\dagger _k \hat{O} \hat{L}_k - \frac{1}{2}\lbrace  
     \hat{L}^\dagger _k \hat{L}_k , \hat{O}
     \rbrace .
 \end{equation}

It is rather straight-forward to see that $\partial_t \hat{N}  =0 $ for the QASIP.

% =============================================================================================================================================================================== % 

\section{Additional numerical data for the Bose-Hubbard model \label{app:BH model}}

Here we present further technical details on the numerical analysis of the Bose-Hubbard model. 

\red{ For large $N$ values, the Von Neumann entanglement entropy is easier to obtain than the logarithmic negativity. In the Appendix \ref{App:Entanglement}, it was shown that the pure state Von Neumann entanglement entropy  can be obtained from the ket state. However, the logarithmic negativity requires finding the spectrum of the partially transposed density matrix. The complexity of handling density matrices is certainly higher than that of handling ket states, hence the lower values that were reached for the logarithmic negativity.}

\begin{figure}
    \centering
    \includegraphics[scale=0.4]{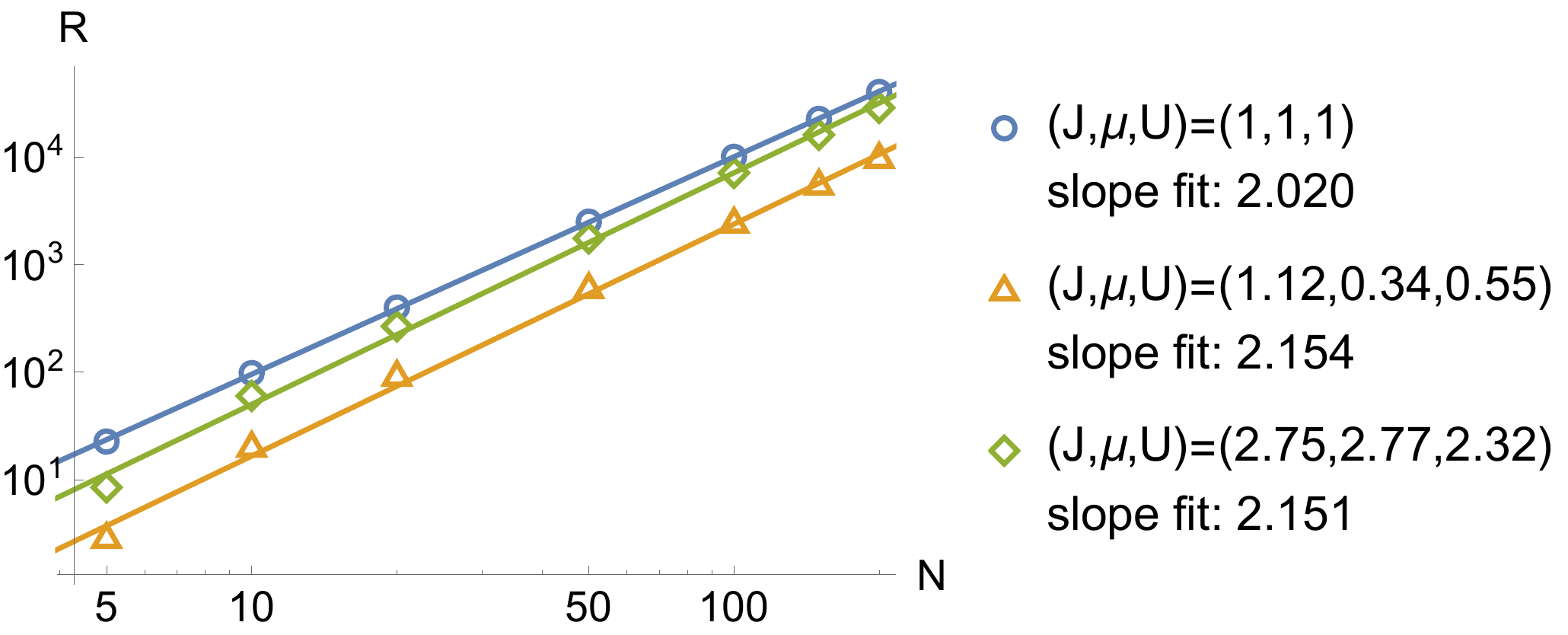}
    \caption{
    \red{In the large $N$ limit of the two-site Bose-Hubbard model, numerical evaluation show that $R \propto N^\beta$ with $\beta \approx 2.15$. The three parameters in the range $N\in \left[1,400\right]$,  are picked to allow for a sufficiently large $R$ values. 
    }}
    \label{fig:BH RofN}
\end{figure}

\red{In Fig.~\ref{fig:BH RofN} the scaling $R \propto N^\beta$ is presented in the large $N$ limit for three values of the parameters $(J,U,\mu)$. The values are picked to produce large $R$ values and to span over a few length scales in $R$, providing a reliable prediction for the exponents.}

\section{Additional numerical data for the QASIP model}

Here we present further technical details on the numerical analysis of the QASIP model.

\begin{figure}
    \centering
    \includegraphics[scale=0.45]{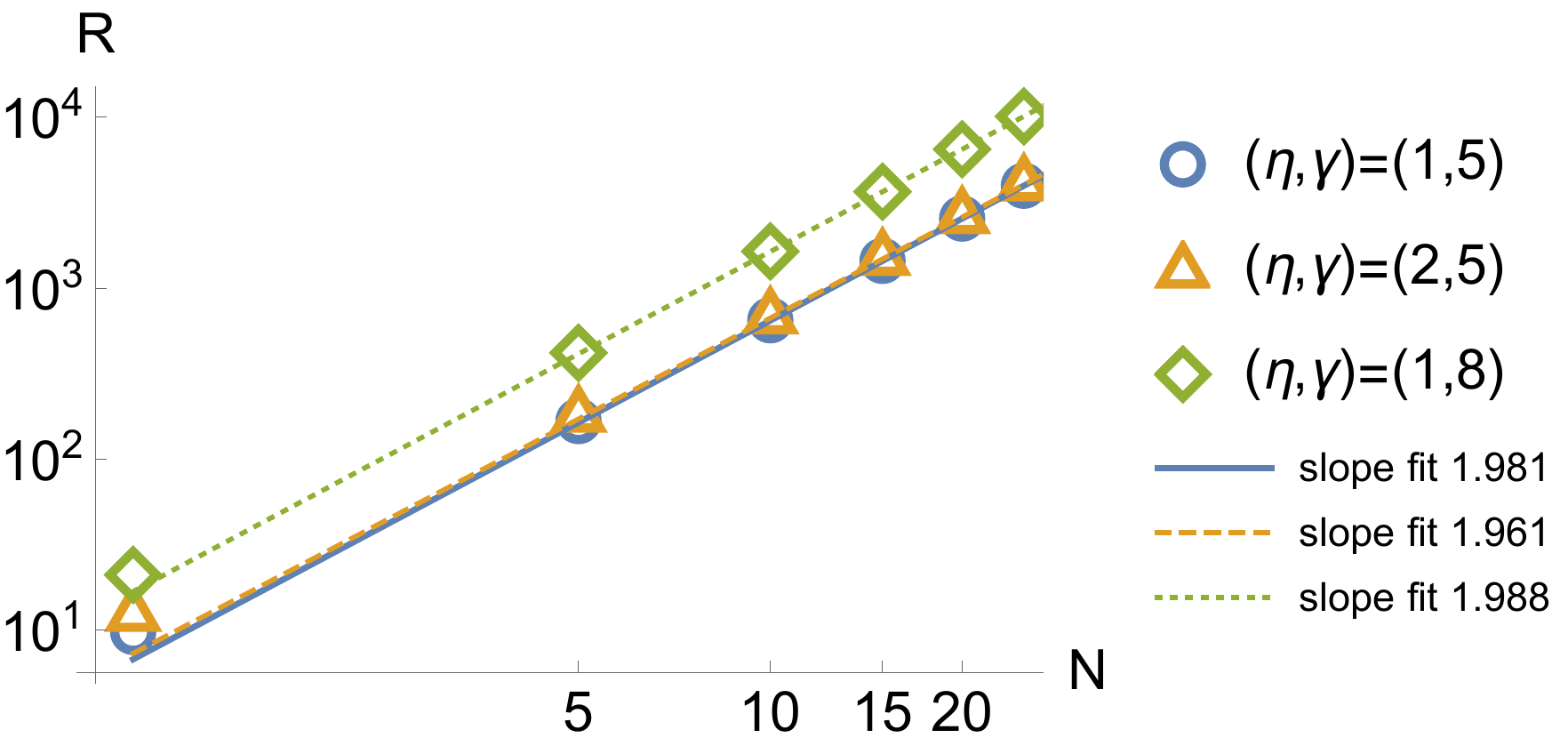}
    \caption{
    \red{In the large $N$ limit of the QASIP model, it is numerically corroborated that $R \propto N^\beta$ where $\beta\approx 1.988$. Different $\eta,\gamma$ values were tested, 
leading to the large $R$ limit at $\varepsilon=1,N\in[1,25]$. 
  }}
    %The scaling of the ratio $R$ with the particle number $N$ for the QASIP,  with  $\varepsilon=\eta  =1  $, and different $\gamma$ values. A close to quadratic scaling is observed.  }
    \label{fig:QASIP R of N scaling }
\end{figure}

In Fig.~\ref{fig:QASIP R of N scaling } the large $N$ scaling of $R\propto N^\beta $ is plotted for different $\gamma$ values. Since testing the logarithmic negativity  in large $N$ values is numerically challenging, the parameters were chosen to facilitate as large $R$ as possible which improves exponent fitting.

\bibliography{Cond_Ent}

\end{document}